\documentstyle[amssymb,amsmath,12pt]{article}

\begin{document}

\title{Worldline Influence Functional: Abraham- Lorentz-Dirac-Langevin Equation
from QED}
\author{Philip R. Johnson\thanks{
Electronic address: {\tt pj19@umail.umd.edu}} and B. L. Hu\thanks{%
Electronic address: {\tt hub@physics.umd.edu}} \\
%EndAName
Department of Physics, University of Maryland\\
College Park, Maryland 20742-4111}
\date{December 1, 2000}
\maketitle

\begin{abstract}
We present a stochastic theory of charges moving in an electromagnetic field
using nonequilibrium quantum field theory. We give a first principles'
derivation of the Abraham-Lorentz-Dirac-Langevin equation which depicts the
quantum expectation value for a particle's trajectory and its stochastic
fluctuations by combining the worldline path integral quantization with the
Feynman-Vernon influence functional or closed-time-path effective action
methods \cite{if,cgea}. At lowest order, the equations of motion are
approximated by a stochastic Lorentz-Dirac equation.
\end{abstract}

\noindent {\small {\it -Talk given by PRJ at the Capri Workshop on Quantum
Aspects of Beam Physics, Oct. 2000. To appear in the Proceedings edited by
Pisin Chen (World Scientific, Singapore, 2001).}}

\section{Introduction}

\subsection{Particle versus field formulation, fixed versus dynamic
background}

The contrasting paradigms of particles versus fields give very different
representations for physical processes. The success of quantum field theory
for the past century has propelled the field concept to the forefront, with
the particle interpreted as excitations of field degrees of freedom. In
recent years, the use of the quantum-mechanical path integral in string
theory has inspired a renewed effort towards particle-centric quantum
formalisms. The so-called worldline quantization method -- where the
particle's spacetime coordinate $x^{\mu }\left( \tau \right) $ is
effectively quantized -- has been used for calculating high-loop processes
in non-dynamical classical background fields \cite{Worldline}. The
background field approximation is useful and well-defined in conditions
where quantum fluctuations of the fields are small. However, many problems
require the inclusion of field dynamics beyond the fixed-background
approximation. A simple example is radiation reaction. More examples are
found in semiclassical gravity, where spacetime playing the role of
`particle' is treated classically in its interaction with quantum fields 
\cite{scg}.

The proper treatment of radiation-reaction requires, at the least,
consistency between the particle's (quantum) average trajectory and the
average radiation. When one treats the particle as quantum mechanical, but
coupled only to the mean-field background (and not the quantum field
fluctuations) one is working within the semiclassical regime \cite
{Kiefer94(Semiclassical.Approx.Q.Gravity)}. (When quantum fluctuations of
the particle motion are sufficiently small, one recovers the classical
regime \cite{Decoherence}). In our program, we go beyond the semiclassical
approximation to include the influence of quantum field fluctuations on the
moving particle \cite{Phil-thesis}, thus entering the stochastic regime. To
do this properly, we must begin with the full quantum theory for both
particle and field, since a particle can never fully decouple from the
field. Also since most field degrees of freedom remain unobserved, we treat
the particle as a quantum open system with the field acting as an
environment \cite{Matacz}. This is how nonequilibrium and stochastic
mechanics ideas enter, and why the self-consistent quantum particle
evolution is generally described by stochastic quantum dynamics \cite{Alpan}.

The open system concept and the accompanying method of coarse-grained
effective action and influence functional have been applied to interacting
quantum fields and semiclassical gravity. The new task we undertake is the
adoption of the worldline quantization method for relativistic charged
particles treated as an open system \cite{Phil-thesis,JH1-3}. This is an
alternative formulation of QED in terms of quantized worldlines {\it and}
quantum fields. The worldline structure is highly efficient for describing
the particle-like degrees of freedom (such as spacetime position), while the
field structure most easily describes processes like radiation.

\subsection{Radiation reaction, causality, and the Abraham-Lorentz-Dirac
equation}

The conceptual and technical scheme we construct is advantageous for
carrying out calculations which conventional methods find difficult, and for
clarifying confusions in existing theories. Consider the classical problem
of radiation reaction. A relativistic solution to classical radiation
reaction is known as the Abraham-Lorentz-Dirac (ALD) equation, whose
solutions give rise to famous paradoxes including: runaway solutions,
pre-acceleration, and the need for higher-derivative initial data \cite
{runaways}. These difficulties with the classical theory may be avoided by
replacing point particles with extended objects \cite
{Moniz.Sharp.Levine.77(extended.charges)}; but a relativistic treatment is
cumbersome. Moniz and Sharp \cite{Moniz.Sharp.Levine.77(extended.charges)}
have shown that non-relativistic QED for extended objects can be causal and
runaway free. Low has shown that runaways apparently don't occur in spin 1/2
QED \cite{Low98:runaways}. Others \cite
{Krivistskii.Tsytovich91(Average.RR.QED)} have shown how the lowest
order-perturbation expansion of the ALD\ equation can be inferred within
conventional relativistic-QED field theory. But most of these earlier works
lack a comprehensive theoretical foundation for addressing all these issues
of quantum field theoretic and nonequilibrium (in fact non-Markovian)
nature, and for dealing with their interrelations consistently and
completely. We show, for instance, that the equations of motion for the
particle's mean trajectory are causal and runaway-free regardless of fine
structures like spin. We highlight the role of the field-environment in
decohering the particle trajectory. Going beyond the result in \cite
{Krivistskii.Tsytovich91(Average.RR.QED)} which is to order $e^{4}$ in the
particle charge, we find the ALD\ equation to all orders in $e$ for the
semiclassical limit. This shows the significant advantage of the worldline
formulation together with the loop expansion in the influence action. The
inclusion of spin and color are important to make full use of this approach
in QED and QCD.

Beyond demonstrating the conceptual and technical consistency of the
particle equations of motion, we describe the influence of quantum field
fluctuations on the particle trajectory. Quantum field fluctuations play
dual roles: they are responsible for the decoherence in the quantum particle
(system), leading to the emergent semiclassical and classical behaviors, and
they provide an effective classical stochastic noise in the particle motion 
\cite{Decoherence}. We show that the one loop equations of motion for the
particle trajectory are approximated by stochastic differential equations
featuring colored noise that encodes the influence of the quantum statistics
of the field . When there is sufficient decoherence, these equations give
excellent approximations to the particle motion. Even in the case of weak
decoherence they provide a one-loop approximation to the particle
correlation functions, and may be extended self-consistently to higher-loop
approximations via the N-particle-irreducible (NPI) effective action \cite
{stubol}. In the second part of this paper, we show the main steps in our
derivation of the ALD-Langevin equations describing the stochastic dynamics
of particle motion.

\section{Coarse-grained effective action and equations of motion}

\subsection{The in-in generating functional}

We begin with the construction of a generating functional for the particle
worldline correlation functions. We assume an initially factorized density
matrix for the particles plus field at initial time $t_{i}$ with the form 
\begin{equation}
\hat{\rho}\left( t_{i}\right) =\hat{\rho}_{particles}\left( t_{i}\right)
\otimes \hat{\rho}_{field}\left( t_{i}\right) \equiv \hat{\rho}_{z}\left(
t_{i}\right) \otimes \hat{\rho}_{\varphi }\left( t_{i}\right) .
\label{initial density matrix}
\end{equation}
For simplicity, consider only one particle worldline and assume that the
initial particle state is a positive frequency configuration space state $%
\hat{\rho}_{z}\left( t_{i}\right) =|z_{i}^{\left( +\right) }\rangle \langle
z_{i}^{\left( +\right) }|,$ where $|z_{i}^{\left( +\right) }\rangle \equiv |%
{\bf z}_{i},t_{i};+\rangle =\hat{\psi}^{\left( +\right) \dagger }\left(
z_{i}\right) |vac\rangle _{particle}.$ These are quasi-localized, physical,
relativistic one-particle states; we treat more general initial states in 
\cite{Phil-thesis,JH1-3}. A basis of field states is $|A\left( x\right)
\rangle ;$ we define a direct product basis for the particles plus field by $%
|zA\rangle =|z\rangle \otimes |A\rangle .$

To find the generating functional for the correlation functions defined as 
\begin{equation}
\langle\hat{z}^{\mu}\left( \tau_{1}\right) ...\hat{z}^{\nu}\left( \tau
_{n}\right) \rangle=\text{Tr}_{\text{Az}}\left( \hat{z}^{\mu}\left(
\tau_{1}\right) ...\hat{z}^{\nu}\left( \tau_{n}\right) \hat{\rho}%
_{in}\right) ,  \label{Correlation functions}
\end{equation}
where $\hat{z}^{\mu}\left( \tau\right) $ are Heisenberg operators and the
trace is over all final particle and field states, we need the time
evolution of the density matrix: 
\begin{equation}
\hat{\rho}\left( t_{f}\right) =\hat{U}\left( t_{f},t_{i}\right) \hat{\rho }%
\left( t_{i}\right) \hat{U}\left( t_{f},t_{i}\right) ^{\dagger}.
\end{equation}
In field theory, one could introduce a functional representation for the
unitary evolution operators $\hat{U}$ in terms of the sum over all
field-histories between $t_{i}$ and $t_{f},$ consistent with the initial and
final state boundary conditions.

Instead, in a hybrid particle-field theory we start with the action
\begin{eqnarray}
S\left[ z,A\right]  &=&S_{z}\left[ z\right] +S_{A}\left[ A\right] +S_{int}%
\left[ z,A\right]   \nonumber \\
&=&\int d\tau \left( \dot{z}^{\mu }\dot{z}_{\mu }/N-Nm^{2}+h_{\mu }\left(
\tau \right) \dot{z}^{\mu }\left( \tau \right) \right) -\int d^{4}x\,F^{2}/4
\nonumber \\
&&-\int d^{4}x\,j^{\mu }\left( x;z\right) \left( A_{\mu }\left( x\right) +%
\bar{A}_{\mu }\left( x\right) \right) .  \label{action}
\end{eqnarray}
The quantum degrees of freedom are the field $A_{\mu }\left( x\right) ,$ the
particle coordinates $z^{\mu }\left( \tau \right) ,$ and the `lapse
function' $N\left( \tau \right) .$ The action $S\left[ z,A\right] $ is
invariant under reparametrizations $\tau \rightarrow \tau ^{\prime }=\tau
+\varepsilon \left( \tau \right) ;$ this is a crucial symmetry of the
relativistic worldline formulation. These issue are discussed in detail in 
\cite{Phil-thesis}. The $F^{2}/4$ term is the usual free electromagnetic
field action. We allow for the possibility of an additional, non-dynamical
background field $\bar{A}.$ The particle couples to the field through the
current $j^{\mu }\left( x;z\right) ,$ which is given by 
\begin{equation}
j^{\mu }\left( x;z\right) =e\int d\tau \dot{z}^{\mu }\left( \tau \right)
\delta \left( x-z\left( \tau \right) \right) .
\end{equation}
In (\ref{action}) we add a source term $h_{\mu }\left( \tau \right) $ whose
role will be made clear below.

The in-in or closed-time-path (CTP) generating functional is given by \cite
{JH1-3}
\begin{eqnarray}
Z_{\text{in-in}}\left[ h,h^{\prime }\right]  &=&\text{Tr}_{Az}\left( \hat{U}%
_{h}\left( t_{f},t_{i}\right) \hat{\rho}_{i}\hat{U}_{h^{\prime }}\left(
t_{f},t_{i}\right) ^{\dagger }\right)   \nonumber \\
&=&\int d{\bf z}_{f}d{\bf z}_{f}^{\prime }dA_{f}dA_{i}dA_{i}^{\prime }\delta
\left( {\bf z}_{f}-{\bf z}_{f}^{\prime }\right) \left( \partial
_{t_{f}}^{\prime }-\partial _{t_{f}}\right)   \nonumber \\
&&\left. \times \langle A_{f}z_{f}|A_{i}z_{i}^{\left( +\right) }\rangle
_{h}\langle A_{f}z_{f}^{\prime }|A_{i}^{\prime }z_{i}^{\left( +\right)
}\rangle _{h^{\prime }}^{\ast }\rho \left( A_{i},A_{i}^{\prime
};t_{i}\right) \right| _{t_{f}=t_{f}^{\prime }}.  \label{Z1}
\end{eqnarray}
We have implicitly assumed in writing (\ref{Z1}) that the sole particle
worldline begins at $z_{i}$ on the initial time hypersurface $\Sigma \left(
t_{i}\right) $ and ends somewhere on the final time hypersurface $\Sigma
\left( t_{f}\right) .$ We then integrate over all possible ${\bf z}_{f}$ on $%
\Sigma \left( t_{f}\right) .$ A more general form, with more general
boundary conditions, is given in \cite{Phil-thesis,JH1-3}. For example, a
worldline that begins and ends on $\Sigma \left( t_{i}\right) $ $\left(
\Sigma \left( t_{f}\right) \right) $ describes pair-annihilation $\left( 
\text{creation}\right) $.

We may now use the worldline formalism to give a path integral
representation for the propagators $\langle A_{f}z_{f}|A_{i}z_{i}^{\left(
+\right) }\rangle _{h}$ and $\langle A_{i}^{\prime }z_{i}^{\left( +\right)
}|A_{f}z_{f}^{\prime }\rangle _{h^{\prime }}=\langle A_{f}z_{f}^{\prime
}|A_{i}^{\prime }z_{i}^{\left( +\right) }\rangle _{h^{\prime }}^{\ast }.$
They are given by 
\begin{equation}
\langle A_{f}z_{f}|A_{i}z_{i}^{\left( +\right) }\rangle
_{h}=\int_{z_{i}A_{i}}^{z_{f}A_{f}}DzDA\exp \left\{ \frac{i}{\hslash }S\left[
z,A,h\right] \right\} ,  \label{U1}
\end{equation}
and similarly for $\langle A_{i}^{\prime }z_{i}^{\left( +\right)
}|A_{f}z_{f}^{\prime }\rangle _{h^{\prime }}.$ The measure is defined by 
\begin{equation}
\int_{z_{i}A_{i}}^{z_{f}A_{f}}DzDA=\prod_{i}\prod_{\mu
,j}\int_{t_{i}}^{t_{f}}dz_{i}^{0}\int_{-\infty }^{\infty }d{\bf z}_{i}\int
dA_{\mu }\left( x_{j}\right) .
\end{equation}
The generating functional then takes the form 
\begin{eqnarray}
Z_{\text{in-in}}\left[ h,h^{\prime }\right] & =\int d{\bf z}%
_{f}\int_{z_{i},z_{i}}^{z_{f},z_{f}^{\prime }=z_{f}}DzDz^{\prime }\left.
\left( \partial _{t_{f}}^{\prime }-\partial _{t_{f}}\right) e^{i\left( S_{z}%
\left[ z,h\right] -S_{z}\left[ z^{\prime },h^{\prime }\right] \right)
}\right| _{t_{f}=t_{f}^{\prime }}  \nonumber \\
& \times \int_{A_{i}A_{i}^{\prime }}^{A_{f}A_{f}^{\prime
}}dA_{f}dA_{i}dA_{i}^{\prime }DA\,DA^{\prime }\rho \left(
A_{i},A_{i}^{\prime };t_{i}\right)  \nonumber \\
& \times \exp \left\{ \frac{i}{\hbar }\left( S_{A}\left[ A\right] +S_{int}%
\left[ z,A\right] -S_{A}\left[ A^{\prime }\right] +S_{int}\left[ z^{\prime
},A^{\prime }\right] \right) \right\} .  \label{Z4}
\end{eqnarray}
The sum over histories in (\ref{U1}) include all worldlines that begin at $%
z_{i},$ end at $z_{f},$ and are bounded between the initial and final times, 
$t_{i}$ and $t_{f},$ respectively, including those that are spacelike and
that change direction in time \cite{Phil-thesis}.

\subsection{Feynman-Vernon influence functional}

Evaluating the generating functional for our action (\ref{action}) is
virtually impossible. The interaction term $S_{int},$ while linear in $A,$
is highly non-linear in the worldline coordinate $z.$ However, if the
initial state of the field $\rho _{\varphi }\left( A_{i},A_{i}^{\prime
};t_{i}\right) $ is Gaussian, the entire integrand in the second line of (%
\ref{Z4}) is Gaussian in the variables $A$. Thus, we can then do the $%
A,A^{\prime }$ path integrals exactly. Initial Gaussian field states include
thermal, squeezed, and coherent states, and therefore provides a fairly rich
set of interesting and physical examples. The result of the $A,A^{\prime }$
functional integrals is the Feynman-Vernon influence functional \cite{if}: 
\begin{eqnarray}
F\left[ z,z^{\prime }\right]  &=&\int_{A_{i}A_{i}^{\prime
}}^{A_{f}A_{f}^{\prime }}dA_{f}dA_{i}dA_{i}^{\prime }DA\,DA^{\prime }\rho
\left( A_{i},A_{i}^{\prime };t_{i}\right)  \\
&&\times \exp \left\{ \frac{i}{\hbar }\left( S_{A}\left[ A\right] +S_{int}%
\left[ z,A\right] -S_{A}\left[ A^{\prime }\right] +S_{int}\left[ z^{\prime
},A^{\prime }\right] \right) \right\} .  \nonumber
\end{eqnarray}
Notice that initial factorizability (see (\ref{initial density matrix}))
makes this result independent of the initial particle state $\hat{\rho}%
_{z}\left( t_{i}\right) .$ The generating functional is now 
\begin{eqnarray}
Z_{\text{in-in}}\left[ h,h^{\prime }\right]  &=&\int d{\bf z}%
_{f}\int_{z_{i}}^{z_{f}}\int_{z_{i}^{\prime }}^{z_{f}^{\prime
}=z_{f}}DzDz^{\prime }\left( \partial _{t_{f}}^{\prime }-\partial
_{t_{f}}\right)   \nonumber \\
&&\times \left. \exp \left\{ \frac{i}{\hslash }\left( S_{z}\left[ z,h\right]
-S_{z}\left[ z^{\prime },h^{\prime }\right] +S_{IF}\left[ z,z^{\prime }%
\right] \right) \right\} \right| _{t_{f}=t_{f}^{\prime }},
\end{eqnarray}
where the influence action $S_{IF}$ is defined via $F\left[ z,z^{\prime }%
\right] =\exp \left( \frac{i}{\hbar }S_{IF}\left[ z,z^{\prime }\right]
\right) .$ For the Gaussian case, the influence action is known; it is 
\begin{eqnarray}
S_{IF} &=&\frac{1}{\hbar }\int \int dxdx^{\prime }g^{\mu \nu }j_{\mu
}^{-}\left( x\right) G_{R}\left( x,x^{\prime }\right) j_{\nu }^{+}\left(
x^{\prime }\right)  \\
&&+\frac{i}{\hbar }\int \int dxdx^{\prime }g^{\mu \nu }\jmath _{\mu
}^{-}\left( x\right) G_{H}\left( x,x^{\prime }\right) j_{\mu }^{-}\left(
x^{\prime }\right) .  \nonumber
\end{eqnarray}
The $j^{\pm }$ are defined by 
\begin{equation}
j_{\mu }^{\pm }\left( x;z,z^{\prime }\right) =\int d\tau \left( \dot{z}_{\mu
}\delta \left( x-z\right) \pm \dot{z}_{\mu }^{\prime }\delta \left(
x-z^{\prime }\right) \right) ,
\end{equation}
$G_{R}\left( x,x^{\prime }\right) $ is the retarded Green's function 
\begin{equation}
G_{R}\left( x,x^{\prime }\right) =\text{Tr}_{A}\left( \left[ \hat{A}_{\mu
}\left( x\right) ,\hat{A}_{\nu }\left( x^{\prime }\right) \right] \hat{\rho}%
_{A}\left( t_{i}\right) \right) \theta \left( t,t^{\prime }\right) ,
\end{equation}
and $G_{H}\left( x,x^{\prime }\right) $ is the Hadamard function 
\begin{equation}
G_{H}\left( x,x^{\prime }\right) =\text{Tr}_{A}\left( \left\{ \hat{A}_{\mu
}\left( x\right) ,\hat{A}_{\nu }\left( x^{\prime }\right) \right\} \hat{\rho}%
_{A}\left( t_{i}\right) \right) .
\end{equation}
The coarse-grained effective action is defined by \cite{cgea} 
\begin{equation}
S_{CGEA}\left[ z^{\pm },h^{\pm }\right] =S_{z}\left[ z,h\right] -S_{z}\left[
z^{\prime },h^{\prime }\right] +S_{IF}\left[ z^{\pm }\right] ,
\end{equation}
where $h_{\mu }^{\pm }\left( \tau \right) =\left( h_{\mu }\left( \tau
\right) \pm h_{\mu }^{\prime }\left( \tau \right) \right) .$

\subsection{Decoherence and the (modified) loop expansion}

To evaluate the non-linear $z,z^{\prime }$ path integrals, we need to use an
approximation method. In perturbative scattering theory, one expands $%
e^{iS_{IF}/\hbar }=1+iS_{IF}/\hbar +{\cal O}\left( e^{4}\right) .$ (Lowest
order interactions are ${\cal O}\left( e^{2}\right) $ because we are working
in a in-in formalism.) But, to evaluate quantum corrections to the
semiclassical result, the loop expansion (or even better, the NPI effective
action \cite{stubol}) is the most suitable. Observe that 
\begin{equation}
\left| F\left[ z^{\pm }\right] \right| =\left| e^{iS_{CGEA}/\hbar }\right|
=e^{-\text{Im}S_{IF}\left[ z^{\pm }\right] }.
\end{equation}
It is not difficult to show that Im$\left( S_{IF}\left[ z^{\pm }\right]
\right) \geq 0$ for all $z^{\pm }.$ For large $z^{-}=z-z^{\prime },$ it is
often the case that $\left| F\right| \ll 1$. This suppression of the
generating functional for large $z^{-}$ is a manifestation of decoherence;
it occurs due to the large number of degrees of freedom $A\left( x\right) $
that are integrated out (traced over) in finding the influence action. We
now assume that decoherence effectively suppresses large $z^{-}$ (to what
extent this is true depends on the details of the system, environment, and
their mutual coupling) and expand $S_{CGEA}$ in powers of $z^{-}.$ We define 
$S_{CGEA}^{R}\equiv $ Re$\left( S_{CGEA}\right) $ and $S_{CGEA}^{I}\equiv $
Im$\left( S_{CGEA}\right) $. Dropping the ${\cal O}\left( z^{-}\right) ^{3}$
terms gives a Gaussian approximation to the $z^{-}$ path integrals. Then the 
$Dz^{-}$ integration may be performed giving 
\begin{eqnarray}
Z_{\text{in-in}}\left[ h,h^{\prime }\right]  &=&\int d{\bf z}%
_{f}\int_{z_{i}^{+}=z_{i}}^{z_{f}^{+}=z_{f}}Dz^{+}%
\int_{z_{i}^{-}=0}^{z_{f}^{-}=0}Dz^{-}\left( \partial _{t_{f}}^{\prime
}-\partial _{t_{f}}\right) e^{i\int d\tau h_{\mu }^{-}z^{+\mu }\left( \tau
\right) } \\
&&\times e^{i\left( \int \left( \frac{\delta S_{CGEA}^{R}}{\delta Z_{\mu
}^{-}}\right) z_{\mu }^{-}d\tau \right) }e^{-\int d\tau d\tau ^{\prime
}g^{\mu \nu }z_{\mu }^{-}\left( \tau \right) G_{H}\left( z^{+}\left( \tau
\right) ,z^{+}\left( \tau ^{\prime }\right) \right) z_{\nu }^{-}\left( \tau
^{\prime }\right) }  \nonumber \\
&=&\int d{\bf z}_{f}\left( \partial _{t_{f}}^{\prime }-\partial
_{t_{f}}\right) \int_{z_{i}^{+}=z_{i}}^{z_{f}^{+}=z_{f}}Dz^{+}  \nonumber \\
&&\times \exp \left\{ \frac{i}{\hslash }\int d\tau h_{\mu }^{-}z^{+\mu
}\left( \tau \right) -\int d\tau d\tau ^{\prime }\eta _{\mu }\left( \tau
\right) G_{H}\left( z^{+},z^{+\prime }\right) \eta ^{\mu }\left( \tau
^{\prime }\right) \right\} ,  \nonumber
\end{eqnarray}
where we have defined the variable $\eta _{\mu }\left( \tau \right) $ by 
\begin{equation}
\eta ^{\mu }\left( \tau \right) =\left. \frac{\delta S_{CGEA}^{R}\left[
z^{+},z^{+}\right] }{\delta Z_{\mu }^{-}}\right| _{z^{-}=0}.
\label{langevin equation}
\end{equation}
We now have only the $z^{+}$ integration left. Note that (\ref{langevin
equation}) is an equation for $z^{+}$ as a functional of $\eta \left( \tau
\right) .$ We use this fact to change integration variables from $%
z^{+}\rightarrow \eta ,$ yielding $Dz^{+}=($Jacobian)$\times D\eta =D\eta ,$
since the Jacobian equals $1.$ This gives us our final result, the
generating function 
\begin{equation}
Z_{\text{in-in}}\left[ h_{\mu }^{-}\right] =\int D\eta P\left[ \eta \right]
\exp \left\{ \frac{i}{\hslash }\int d\tau \,h_{\mu }^{-}\left( \tau \right)
\,z^{\mu +}\left( \tau \right) \right\} 
\end{equation}
with 
\begin{equation}
P\left[ \eta \right] =\exp \left\{ -\int d\tau d\tau ^{\prime }\eta _{\mu
}\left( \tau \right) G_{H}\left( z^{+},z^{+\prime }\right) \eta ^{\mu
}\left( \tau ^{\prime }\right) \right\} .
\end{equation}
$P\left[ \eta \right] $ may now be treated as a probability distribution for 
$\eta \left( \tau \right) $, thus rendering $\eta ^{\mu }\left( \tau \right) 
$ a stochastic noise variable. (\ref{langevin equation}) is therefore a
Langevin equation for $z_{\mu }^{+}\left( \tau \right) $ with noise $\eta
^{\mu }\left( \tau \right) .$ Derivatives $\delta ^{n}Z_{\text{in-in}}\left[
h_{\mu }^{-}\right] /\delta h_{\mu }^{-}\left( \tau _{1}\right) ...\delta
h_{\nu }^{-}\left( \tau _{n}\right) $ give the correlation functions (\ref
{Correlation functions}). Because $z^{+}$ is a nonlinear function of $\eta ,$
as determined by (\ref{langevin equation}), and $G_{H}\left(
z^{+},z^{+\prime }\right) $ is a function of $z^{+},$ the distribution $P%
\left[ \eta \right] $ is {\it not }Gaussian. In \cite{JH1-3}, we show how to
perform a cumulant expansion on the Langevin equation. The order $e^{2}$
term then generates Gaussian noise, but more generally, one recognizes that
even after decoherence has suppressed large quantum fluctuations in $z^{-},$
the effective noise will be non-Gaussian for nonlinear Langevin equations.
Only when the Langevin equation is linear is the noise strictly Gaussian; at
high temperatures it is approximately white.

\section{The ALD-Langevin equation}

To apply this general result to charged particles in the quantum
electromagnetic field requires regularizing $S_{CGEA}$ by introducing a
high-frequency cutoff $\Lambda $ in the Green's functions $G_{R,H}^{\Lambda
}.$ We perform this analysis in \cite{JH1-3}. The resulting Langevin
equations of motion found from (\ref{langevin equation}) are 
\begin{equation}
m\left( r\right) \ddot{z}_{\mu }\left( \tau \right) =e\dot{z}^{\mu }F_{\mu
\nu }^{ext}+e^{2}\sum_{n=2}^{\infty }g^{\left( n\right) }(r)u_{\mu }^{\left(
n\right) }+\eta _{\mu }\left( \tau \right) .  \label{ALDL}
\end{equation}
$F_{\mu \nu }^{ext}$ is included as a possible classical, external
background field. The variable $r$ $=\tau -\tau _{i}$ is the elapsed
proper-time from the initial time $\tau _{i}$.

Quantum field fluctuations dress the particle state giving it an effective
mass $m\left( r\right) .$ At $r=0$ (corresponding to the initial time when
the field and particle are uncorrelated) $m\left( 0\right) =m_{0},$ the bare
mass of the particle. The mass is then renormalized on the cutoff timescale $%
1/\Lambda,$ quickly reaching its late-time value $m\left( \infty\right)
=m_{0}+a_{0}\Lambda.$ The constant $a_{0}$ depends on the details of the
cutoff since the high-energy theory ultimately determines the precise
relationship between bare and effective mass.

Radiation reaction is given by the second term on the RHS of (\ref{ALDL}).
The 
\begin{equation}
u_{\mu }^{\left( n\right) }=u_{\mu }^{\left( n\right) }\left( \acute{z},%
\ddot{z},...,d^{n+1}z/d\tau ^{n}\right)
\end{equation}
are combinations of higher time-derivatives of $z\left( \tau \right) $. The
time dependent coefficients $g^{\left( n\right) }\left( r\right) $
determined by the regulated retarded Greens function $G_{R}^{\Lambda }$
scale as $\Lambda ^{2-n}$ with the cutoff. Therefore, the $n>2$ terms are
high-energy corrections to the radiation reaction force and as such they are
strongly suppressed at low-energies. The late-time behavior of the $n=2$
term gives 
\begin{equation}
f_{\mu }^{R.R.\left( 2\right) }=\frac{e^{2}}{2\pi }\left( \dot{z}_{\mu }%
\ddot{z}^{2}+\stackrel{...}{z}_{\mu }\right) ,
\end{equation}
which is just the usual ALD results. In \cite{JH1-3}, the expansion (\ref
{ALDL}) is shown to be convergent for all $r,$ including $r\rightarrow
\infty $ and $r\rightarrow 0.$ The short-time behavior is particularly
significant: \thinspace $\lim_{r\rightarrow 0}g^{\left( n\right) }\left(
r\right) =0$ for all $n;$ therefore, the radiation reaction force
identically vanishes at $t_{i},$ and the usual Newtonian initial data (i.e.
position and velocity) uniquely determine the equations of motion. In \cite
{JH1-3}, it is further shown that the equations of motion are strictly
causal and without runaways. The non-Markovian time-dependence of the
radiation reaction force reflects the fact that in any nonequilibrium
quantum setting it takes time for the particle's self-field to adjust to
changes in its motion; it can not respond faster than the effective
high-energy cutoff $\Lambda .$ This lag is always enough to preserve
causality. To the extent that QED is a good effective field theory, these
results demonstrate the consistency of the low-energy physics, including the
ALD equation as the proper low-energy limit for radiation reaction.

Finally, the noise is given by 
\begin{equation}
\langle \left\{ \eta _{\mu }\left( \tau \right) ,\eta _{\nu }\left( \tau
\right) \right\} \rangle _{s}=g_{\mu \nu }G_{H}^{\Lambda }\left( \tau ,\tau
^{\prime }\right) ,  \label{noise}
\end{equation}
where $\langle \rangle _{s}$ is the noise average with respect to the
distribution $P\left[ \eta \right] .$ Consistency with the
fluctuation-dissipation relation derived in \cite{JH1-3} requires mutually
consistent regularization of both $G_{R}$ and $G_{H}.$ Therefore, the noise
will also depend on the cutoff $\Lambda .$ In fact, the noise kernel is
modified at early times by the restriction that the path integrals over $%
z^{\pm }$ are bounded between the surfaces $\Sigma (t_{i})$ and $\Sigma
\left( t_{f}\right) .$ This modification of the noise is balanced by the
early time radiation reaction behavior. At late times, the restriction on
the paths has little effect on the integrals and the noise quickly
approaches the form (\ref{noise}) just as the radiation reaction force
approaches its late-time limit.\newline

\section*{Acknowledgment}

We thank Pisin Chen for his invitation to this interesting workshop and
Stefania Petracca for her warm hospitality. This research is supported in
part by NSF grant PHY98-00967 and DOE grant DEFG0296ER40949.

\end{document}